\documentclass[12]{iopart}
\usepackage{graphicx}

\begin{document}

\title{Coexistence of SDW, d-wave singlet and staggered $\pi$-triplet superconductivity}
\author{A Aperis$^1$, G Varelogiannis$^2$, P B Littlewood$^3$ and B D Simons$^3$}
\address{$^{1,2}$ Department of Physics, National Technical University
of Athens, GR-15780 Athens, Greece}
\address{$^{^3}$ Cavendish Laboratory, University of Cambridge, Cambridge CB3 03E, United Kingdom}

\eads{\mailto{$^1$aaperis@central.ntua.gr}, \mailto{$^2$varelogi@central.ntua.gr}}

\begin{abstract}
We have studied the competition and coexistence of staggered triplet
SC with d-wave singlet SC and SDW in the mean-field approximation.
Detailed numerical studies demonstrate that particle-hole asymmetry
mixes these states and therefore they are simultaneously present.
Even more interesting were the results of our study of the influence
of a uniform magnetic field. We observe novel transitions that show
the characteristics of Fulde-Ferrel phases, yet they concern
transitions to different combinations of the above orders.
For example, above a given field, in a particle-hole symmetric
system we observe a transition from d-wave singlet SC to a
state in which d-wave singlet SC coexists with staggered triplet
SC and SDW. We believe our results may provide, among others,
a direct explanation to recent puzzles about the Fulde Ferrel
like states that are apparently observed in CeCoIn5.
\end{abstract}

\newcommand{\bm}{\mathbf}
\newcommand{\bt}{\textbf}
\newcommand{\mb}{\mathbf}
\newcommand{\phd}{\phantom{\dag}}
\newcommand{\ph}{\phantom{.}}
\newcommand{\noi}{\noindent}
\newcommand{\no}{\nonumber}

\maketitle

\section{Introduction}

Some of the most intriguing research problems in the field of
strongly correlated electrons concern the coexistence of
superconductivity (SC) with itinerant antiferromagnetism, namely the
Spin Density Wave (SDW) state\cite{Overhauser1,Overhauser2,Gruner}. Most of the
electronic systems that have been proposed to exhibit such a
coexisting phase, are thought to be unconventional
superconductors\cite{Sigrist}. The quasi one-dimensional organic
salts based on TMTSF were the the first to exhibit clearly the
proximity of the SDW state with SC \cite{jerome}. More recent
experiments on the same salts have established that there is a
region of macroscopic coexistence of these two phases
\cite{vuletic}. Similar phenomena have also been oberved in organic
quasi-2D superconductors like $\kappa-(ET)_2Cu(NCS)_2$, High-$T_c$
oxides in the underdoped regime and heavy-fermion compounds such as
Uranium based $UBe_{13}$, $UPt_3$ or Cerium based $CeRhIn_5$ just to
name a few\cite{Gabovich}. Many controversies still remain about the
type of SC involved, and in particular whether SC is singlet or
triplet.

In the present paper we study the coexistence of singlet d-wave SC
with SDW and in particular we focus on the induced staggered triplet
SC component and new behavior that may result because of the
presence of these \textit{three} order parameters. Several studies
of the competition of unconventional SC with SDW and the possibility
of an emergent staggered $\pi-triplet$ pairing are available
\cite{Machida0,Machida1,Psaltakis,Maitra,Murakami,Kyung}. For
example, in \cite{Psaltakis,Murakami} scattering by non-magnetic
impurities is shown to stabilize a phase with all three Order
Parameters (OPs) whereas in \cite{Maitra,Kyung} the induced
$\pi$-pairing is studied in the context of deviations from
half-filling.

We associate here for the first time the coexistence and competition
of the above triplet of phases with the recent experimental evidence
of a Fulde Ferrel Larkin Ovchinikov (FFLO) \cite{FFLO,Larkin} state
in the high field-low temperature phase diagram regime of
$CeCoIn_5$\cite{Correa,mats}. This quasi-2D compound has remarkable
features separating it from other HF superconductors, such as the
highest SC $T_c$ among all Ce and U compounds. It's gap symmetry is
considered to be d-wave and it's fermi surface consists of
quasi-cylindrical sheets (for a recent review, see \cite{Matsuda}).
There is no clear evidence up to now for AFM ordering in $CeCoIn_5$.
Nevertheless, it's proximity to $CeRhIn_5$, which is an AFM
uncoventional superconductor\cite{Gabovich}, and recent NMR
experiments suggest that $CeCoIn_5$ is close to a magnetic quantum
critical point situated at a slightly negative
pressure\cite{Miclea}.

Using a eight component spinor formalism and a mean-field approach
we study the influence of varying temperature, doping and external
magnetic field on the competition of d-wave singlet SC with SDW and
staggered $\pi-triplet$ SC. We have considered only the Zeeman
splitting of an in-plane magnetic field which is the relevant term
in the context of a FFLO formulation in a thin film superconductor
whose thickness is smaller than its coherence length. In such a case
orbital effects can safely be neglected. The layered $CeCoIn_5$
meets well these requirements. Bearing in mind the above, we show
how particle-hole asymmetry mixes the above states as proposed
before in slightly different contexts. Most importantly, we produce
novel field-induced transitions that bear remarkable similarities
with the experimentally observed FFLO phases.

\section{Model-Formalism}

Our starting point is the generalized mean-field hamiltonian which
corresponds to the coexistence of d-wave singlet SC,
 $\pi$-triplet (or \textbf{Q}-triplet) SC and SDW order
 parameters under the influence of an external magnetic field:
\begin{eqnarray}\no\fl
{\cal H}=\sum_{\bi{k},\alpha}\xi_{\bi{k}\alpha}c_{\bi{k}
\alpha}^\dagger c_{\bi{k}\alpha}-\sum_{\bi{k},\alpha,\beta}
(\sigma\cdot\bi{n})_{\alpha\beta}M_{\bi{k}}
\Bigl(c_{\bi{k}\alpha}^{\dagger}c_{\bi{k}+\bi{Q}\beta}+hc\Bigl)\\\no
-\frac{1}{2}\sum_{\bi{k},\alpha,\beta}(\rmi\hat{\sigma}_2)_{\alpha\beta}
\Bigl(\Delta_{\bi{k}}c_{\bi{k}\alpha}^{\dagger}c_{-\bi{k}\beta}^{\dagger}+hc
\Bigl)\\\no
-\frac{1}{2}\sum_{\bi{k},\alpha,\beta}\rmi\hat{\sigma}_2(\sigma\cdot\bi{n})_{\alpha\beta}
\Bigl(\Pi_{\bi{k}}^{-\bi{Q}}c_{-\bi{k}-\bi{Q}\alpha}^{\dagger}c_{\bi{k}\beta}^{\dagger}+
\Pi_{\bi{k}}^{\bi{Q}}c_{\bi{k}+\bi{Q}\alpha}^{\dagger}c_{-\bi{k}\beta}^{\dagger}+hc\Bigl)\\
-\mu_{{\rm B}}H\sum_{\bi{k},\alpha,\beta}(\sigma\cdot\bi{n})_{\alpha\beta}
\Bigl(c_{\bi{k}\alpha}^{\dagger}c_{\bi{k}\beta}+hc\Bigl)\end{eqnarray}
where $\alpha,\beta$ are spin indices, $M_{\bi{k}}$,$\Delta_{\bi{k}}$ and
$\Pi_{\bi{k}}$ are the SDW, the d-wave singlet SC and the $\pi$-triplet
SC order parameters respectively, $\mu_{{\rm B}}H$ is the Zeeman term
for a static uniform field and {$\bi n$} is the polarization of
the SDW which is taken \textit{parallel} to that of the $\pi$-triplet
SC spins and the external magnetic field. We have chosen the
z-axis components for our calculations.

For the odd in momentum spin triplet SC order parameter (OP) we have
$\Pi_{\bi{k}}^{-\bi{Q}}=-\Pi_{\bi{k}}^{\bi{Q}}=\Pi_{\bi{k}}$ and
$\Pi_{\bi{k}}^*=\Pi_{\bi{k}}$. Note that the triplet SC component
that we consider is  \textit{staggered}, meaning that the SC pairs
have a finite momentum. Similar $\pi$ operators were introduced in
the past, i.e. by Yang Sun et al. in an SU(4) model for HTC
superconductivity\cite{YangSun}, but, to our knowledge were first
discussed by Psaltakis \etal\cite{Psaltakis}.

In the two-dimensional tetragonal system that we consider here, the
transformations with respect to ${\bi Q}=(\pi,\pi)$ are fundamental.
The wavevector ${\bi Q}$ is commensuratemeaning that translations
from $\bi{k}$ to $\bi{k}+2\bi{Q}$ brings us back at the same place
of the BZ. The electronic dispersion can be generically decomposed
into periodic and antiperiodic terms with respect to $\bi{Q}$:
$\xi_{\bi{k}}=\gamma_{\bi{k}}+\delta_{\bi{k}}$ where
$\gamma_{\bi{k}+\bi{Q}}=-\gamma_{\bi{k}}$ and
$\delta_{\bi{k}+\bi{Q}}= \delta_{\bi{k}}$. Here, as an example, we
assume a single band tight-binding dispersion where
$\gamma_{\bi{k}}=-t_1(\cos k_x+\cos k_y)$ and
$\delta_{\bi{k}}=-t_2\cos k_x\cos k_y$. For $\delta_{{\bi k}}=0$ the
system is particle-hole symmetric and perfectly nested at the
wavevector \textbf{Q}. When $\delta_{\bi k}<0$($\delta_{\bi k}>0$)
the system is considered  e(h)-doped and deviates from perfect
nesting.

In order to treat all order parameters on the same footing we
adopt an eight component spinor space formalism
defined by the following spinor\cite{Psaltakis,Nass}:
\begin{equation}
\Psi^{\dagger}_{\bi{k}}=\bigl(
c^{\dagger}_{\bi{k}\uparrow},c^{\dagger}_{\bi{k}\downarrow},
c_{-\bi{k}\uparrow},c_{-\bi{k}\downarrow},
c^{\dagger}_{\bi{k}+\bi{Q}\uparrow},c^{\dagger}_{\bi{k}+\bi{Q}\downarrow},
c_{-\bi{k}-\bi{Q}\uparrow},c_{-\bi{k}-\bi{Q}\downarrow}\bigr)
\end{equation}
the following tensor products
\begin{eqnarray}
\widehat{\tau}=\widehat{\sigma}\otimes\bigl(\widehat{I}\otimes
\widehat{I})\hskip 0.5cm
\widehat{\rho}=\widehat{I}\otimes\bigl(\widehat{\sigma}\otimes
\widehat{I})\hskip 0.5cm
\widehat{\sigma}=\widehat{I}\otimes\bigl(\widehat{I}\otimes
\widehat{\sigma}) \end{eqnarray}
with $\widehat{\sigma}$ being the
Pauli matrices, form a convenient basis for the projection of the
hamiltonian in this eight component spinor space.

In the above formalism and assuming all OPs real,
our mean-field hamiltonian is rewritten in the compact form:
\begin{equation}
{\cal H}=\sum_{\bi{k}}\Psi_{\bi{k}}^{\dagger}\hat{E}_{\bi{k}}\Psi_{\bi{k}}
\end{equation}
where
\begin{equation}\fl
\hat{E}_{\bi{k}}=\gamma_{\bi{k}}\hat{\tau}_3\hat{\rho}_3+
\delta_{\bi{k}}\hat{\rho}_3-M_{\bi{k}}\hat{\tau}_1\hat{\rho}_3\hat{\sigma}_3+
\Pi_{\bi{k}}\hat{\tau}_2\hat{\rho}_2\hat{\sigma}_1+
\Delta_{\bi{k}}\hat{\tau}_3\hat{\rho}_2\hat{\sigma}_2-\mu_{{\rm B}}H\hat{\rho}_3\hat{\sigma}_3
\end{equation}
is the matrix of the system eigenenergies. The corresponding propagator reads:
\begin{eqnarray}\nonumber\fl
\hat{{\cal G}}_o({\bi{k}},\rmi\omega_n)=\left(-\rmi\omega_n-\hat{E}_{\bi{k}}\right)
\otimes\bigl[A({\bi{k}}',\omega_n)\hat{\tau}_2+
2\bigl(i\gamma_{\bi{k}}\left(\delta_{\bi{k}}-H\mu_{{\rm B}} \hat{\sigma}_3\right)\hat{\tau}_1\\\nonumber
+\bigl(\delta_{\bi{k}}^2+H^2\mu_{{\rm B}}^2
-H\delta_{\bi{k}}\mu_{{\rm B}} \hat{\sigma}_3\bigl)\hat{\tau}_2
+i\left(-HM_{\bi{k}}\mu_{{\rm B}}+\left(M_{\bi{k}}\delta_{\bi{k}}+
\Delta_{\bi{k}}\Pi_{\bi{k}}\right)\hat{\sigma}_3\right)\hat{\tau}_3\\\nonumber
-\hat{\rho}_1\left(H\mu_{{\rm B}}\Pi_{\bi{k}} \hat{\sigma}_2-
i\hat{\sigma}_1\left(H\Delta_{\bi{k}}\mu_{{\rm B}} \hat{\tau}_1+
\gamma_{\bi{k}}\Pi_{\bi{k}} \hat{\tau}_3\right)\right)\bigl)\bigl]\otimes\\\nonumber
\bigl[\left(B({\bi{k}}',\omega_n)+8H\mu_{{\rm B}}\left(-M_{\bi{k}}
\Delta_{\bi{k}}\Pi_{\bi{k}}+\delta_{\bi{k}}\left(\Delta_{\bi{k}}^2+
\Pi_{\bi{k}}^2+\omega^2\right)\right)\hat{\sigma}_3\right)\hat{\tau}_2\\\nonumber
-8H\mu_{{\rm B}}\hat{\rho}_1\bigl(\hat{\sigma}_2\bigl(M_{\bi{k}}
\delta_{\bi{k}}\Delta_{\bi{k}}+\left(\gamma_{\bi{k}}^2-\delta_{\bi{k}}^2+
\Delta_{\bi{k}}^2-H^2\mu_{{\rm B}}^2\right)\Pi_{\bi{k}}\\\nonumber
-H\mu_{{\rm B}}\left(\delta_{\bi{k}}\Delta_{\bi{k}}-M_{\bi{k}}
\Pi_{\bi{k}}\right)\hat{\tau}_1\bigl)-\gamma_{\bi{k}}
\left(\delta_{\bi{k}}\Delta_{\bi{k}}-M_{\bi{k}}\Pi_{\bi{k}}\right)\hat{\sigma}_1\hat{\tau}_2\bigl)\\
-4A({\bi{k}}',\omega_n)H\mu_{{\rm B}}\left(\delta_{\bi{k}}
\hat{\sigma}_3\hat{\tau}_2-\Pi_{\bi{k}} \hat{\rho}_1\hat{\sigma}_2\right)
\bigl]\times D({\bi{k}}',\omega_n)
\end{eqnarray}
where
\begin{eqnarray}\nonumber\fl
A({\bi{k}}',\omega_n)=M_{{\bi{k}}'}^2+\gamma_{{\bi{k}}'}^2-
\delta_{{\bi{k}}'}^2+\Delta_{{\bi{k}}'}^2-H^2\mu_{{\rm B}}^2+
\Pi_{{\bi{k}}'}^2+\omega_n^2\\\nonumber\fl
B({\bi{k}}',\omega_n)=A({\bi{k}}',\omega_n)^2-4\Bigl(2M_{\bi{k}}
\delta_{\bi{k}}\Delta_{\bi{k}}\Pi_{\bi{k}}+\left(\gamma_{\bi{k}}^2+
\Delta_{\bi{k}}^2-2H^2\mu_{{\rm B}}^2\right)\Pi_{\bi{k}}^2\\\nonumber
-H^2\mu_{{\rm B}}^2\omega_n^2-\delta_{\bi{k}}^2\left(\Delta_{\bi{k}}^2+
H^2\mu_{{\rm B}}^2+\Pi_{\bi{k}}^2+\omega_n^2\right)\Bigl)\\\no\fl
D({\bi{k}}',\omega_n)=\left[\left(\omega_n^2+E_{++}^2({\bi{k}}')\right)
\left(\omega_n^2+E_{+-}^2({\bi{k}}')\right)\left(\omega_n^2+E_{-+}^2({\bi{k}}')
\right)\left(\omega_n^2+E_{--}^2({\bi{k}}')\right)\right]^{-1}
\end{eqnarray}\\

The poles of the Green function are the following:
\begin{eqnarray}
E_{+\pm}({\bi{k}})&=\mu_{{\rm B}}H+\sqrt{M_{\bi{k}}^2+
\gamma_{\bi{k}}^2+\delta_{\bi{k}}^2+\Delta_{\bi{k}}^2+\Pi_{\bi{k}}^2\pm\Gamma({\bi{k}})}\\
E_{-\pm}({\bi{k}})&=\mu_{{\rm B}}H-\sqrt{M_{\bi{k}}^2+
\gamma_{\bi{k}}^2+\delta_{\bi{k}}^2+\Delta_{\bi{k}}^2+
\Pi_{\bi{k}}^2\pm\Gamma({\bi{k}})}\\\nonumber
\Gamma({\bi{k}})&=2\sqrt{\left(M_{\bi{k}}^2+\gamma_{\bi{k}}^2\right)
\delta_{\bi{k}}^2+2\delta_{\bi{k}} M_{\bi{k}}\Delta_{\bi{k}}\Pi_{\bi{k}}+
\left(\gamma_{\bi{k}}^2+\Delta_{\bi{k}}^2\right)\Pi_{\bi{k}}^2}
\end{eqnarray}
From the above we see that the $\delta_{\bi{k}}$ term, when
non-zero, induces additional terms that may help in the formation of
new fermi sheets when one of the branches goes to zero.

After projecting the above propagator on the different
particle-hole and particle-particle channels, we arrive at
the following system of coupled self-consistent equations obeyed by the three OPs:
\begin{eqnarray}\no\fl
M_{\bi{k}}=T\sum_{{\bi{k}}'}\sum_nV_{{\bi{k}\bi{k}'}}^{SDW}\times
\Bigg\{M_{{\bi{k}}'}\biggl(C({\bi{k}}',\omega_n)+2A({\bi{k}}',\omega_n)^2
\delta_{{\bi{k}}'}^2\\\no
-4A({\bi{k}}',\omega_n)\delta_{{\bi{k}}'}^2(M_{{\bi{k}}'}^2+\gamma_{{\bi{k}}'}^2-
\delta_{{\bi{k}}'}^2)+16\delta_{{\bi{k}}'}^2\left(\Delta_{{\bi{k}}'}^2\Pi_{{\bi{k}}'}^2+
\mu_{{\rm B}}^2H^2\omega_n^2\right)\biggl)\\\no
-2\delta_{{\bi{k}}'}\Delta_{{\bi{k}}'}\Pi_{{\bi{k}}'}
\bigl[A({\bi{k}}',\omega_n)^2-
f4\bigl((\gamma_{{\bi{k}}'}^2+\Delta_{{\bi{k}}'}^2)\Pi_{{\bi{k}}'}^2+
\mu_{{\rm B}}^2H^2\omega_n^2\\
-\delta_{{\bi{k}}'}^2(\Delta_{{\bi{k}}'}^2-\mu_{{\rm B}}^2H^2+\Pi_{{\bi{k}}'}^2+
\omega_n^2)\bigl)\bigl]\Biggl\}\times D({\bi{k}}',\omega_n)\\\no\fl
\Delta_{\bi{k}}=T\sum_{{\bi{k}}'}\sum_nV_{{\bi{k}\bi{k}'}}^{dSC}
\Bigg\{\Delta_{{\bi{k}}'}\biggl(-C({\bi{k}}',\omega_n)+ 2A({\bi{k}}',\omega_n)^2\Pi_{{\bi{k}}'}^2\\\no
-4A({\bi{k}}',\omega_n)\delta_{{\bi{k}}'}^2\left(\Delta_{{\bi{k}}'}^2-
H^2\mu_{{\rm B}}^2+5\Pi_{{\bi{k}}'}^2+\omega_n^2\right)+
8\bigl[3 M_{{\bi{k}}'} \delta_{{\bi{k}}'} ^3 \Delta_{{\bi{k}}'}  \Pi_{{\bi{k}}'}\\\no
-3M_{{\bi{k}}'}\delta_{{\bi{k}}'}\Delta_{{\bi{k}}'}\Pi_{{\bi{k}}'}^3-
\left(\gamma_{{\bi{k}}'} ^2+\Delta_{{\bi{k}}'} ^2\right) \Pi_{{\bi{k}}'}^4+
\delta_{{\bi{k}}'} ^2 \Pi_{{\bi{k}}'} ^2 (3 \gamma_{{\bi{k}}'} ^2+
4 \Delta_{{\bi{k}}'}^2\\\no-3 H^2\mu_{{\rm B}}^2+3 \Pi_{{\bi{k}}'} ^2)-
\delta_{{\bi{k}}'} ^4(\Delta_{{\bi{k}}'} ^2-H^2 \mu_{{\rm B}} ^2+3 \Pi_{{\bi{k}}'} ^2)-
\bigl(\delta_{{\bi{k}}'} ^4+H^2 \mu_{{\rm B}} ^2 \Pi_{{\bi{k}}'} ^2\\\no
+\delta_{{\bi{k}}'} ^2\bigl(H^2 \mu_{{\rm B}} ^2-3\Pi_{{\bi{k}}'} ^2\bigl)\bigl)
\omega_n ^2\bigl]\biggl)+2\delta_{{\bi{k}}'} M_{{\bi{k}}'}
\Pi_{{\bi{k}}'}\bigl[A({\bi{k}}',\omega_n)^2-4
\bigl(\gamma_{{\bi{k}}'} ^2 \Pi_{{\bi{k}}'} ^2\\
+H^2 \mu_{{\rm B}} ^2 \left(\delta_{{\bi{k}}'} ^2+
\omega_n ^2\right)-\delta_{{\bi{k}}'} ^2 \left(\Pi_{{\bi{k}}'} ^2+
\omega_n ^2\right)\bigl)\bigl]\Biggl\}\times D({\bi{k}}',\omega_n)\\\no\fl
\Pi_{\bi{k}}=T\sum_{{\bi{k}}'}\sum_nV_{{\bi{k}\bi{k}'}}^{Qtr}
\Bigg\{-\Pi_{{\bi{k}}'}\biggl(C({\bi{k}}',\omega_n)-2A({\bi{k}}',\omega_n)^2
\left(\gamma_{{\bi{k}}'}^2+\Delta_{{\bi{k}}'}^2\right)+4A({\bi{k}}',\omega_n)
\delta_{{\bi{k}}'}^2\\\no
\cdot\left(5\Delta_{{\bi{k}}'}^2-H^2\mu_{{\rm B}}^2+\Pi_{{\bi{k}}'}^2+
\omega_n^2\right)+8\bigl[-3M_{{\bi{k}}'}\delta_{{\bi{k}}'}^3\Delta_{{\bi{k}}'}
\Pi_{{\bi{k}}'}+3M_{{\bi{k}}'}\delta_{{\bi{k}}'}\Delta_{{\bi{k}}'}^3\Pi_{{\bi{k}}'}\\\no
+\gamma_{{\bi{k}}'}^4\Pi_{{\bi{k}}'}^2+\Delta_{{\bi{k}}'}^4\Pi_{{\bi{k}}'}^2+
H^2\Delta_{{\bi{k}}'}^2\mu_{{\rm B}}^2\omega_n^2+
\delta_{{\bi{k}}'}^4\left(3\Delta_{{\bi{k}}'}^2-
H^2\mu_{{\rm B}}^2+\Pi_{{\bi{k}}'}^2+\omega_n^2\right)\\\no
+\delta_{{\bi{k}}'}^2\left(-3\Delta_{{\bi{k}}'}^4+
H^2\mu_{{\rm B}}^2\omega_n^2+\Delta_{{\bi{k}}'}^2\left(3H^2\mu_{{\rm B}}^2-
4\Pi_{{\bi{k}}'}^2-3\omega_n^2\right)\right)\\\no
+\gamma_{{\bi{k}}'}^2\bigl(3M_{{\bi{k}}'}\delta_{{\bi{k}}'}
\Delta_{{\bi{k}}'}\Pi_{{\bi{k}}'}+2\Delta_{{\bi{k}}'}^2\Pi_{{\bi{k}}'}^2+
H^2\mu_{{\rm B}}^2\omega_n^2-\delta_{{\bi{k}}'}^2\bigl(3\Delta_{{\bi{k}}'}^2-
H^2\mu_{{\rm B}}^2\\\no
+2\Pi_{{\bi{k}}'}^2+\omega_n^2\bigl)\bigl)\bigl]\biggl)+2M_{{\bi{k}}'}
\delta_{{\bi{k}}'}\Delta_{{\bi{k}}'}\bigl(A({\bi{k}}',\omega_n)^2-4H^2\mu_{{\rm B}}^2\omega_n^2\\
+4\delta_{{\bi{k}}'}^2\left(\Delta_{{\bi{k}}'}^2-H^2\mu_{{\rm B}}^2+
\omega_n^2\right)\bigl)\Biggl\}\times D({\bi{k}}',\omega_n)
\end{eqnarray}
where
\begin{eqnarray}\nonumber\fl
C({\bi{k}}',\omega_n)=A({\bi{k}}',\omega_n)\Bigl(A({\bi{k}}',\omega_n)^2+
2A({\bi{k}}',\omega_n)\delta_{{\bi{k}}'}^2\\\nonumber
-8\delta_{{\bi{k}}'}M_{{\bi{k}}'}\Delta_{{\bi{k}}'}\Pi_{{\bi{k}}'}-
4\left(\gamma_{{\bi{k}}'}^2+\Delta_{{\bi{k}}'}^2\right)
\Pi_{{\bi{k}}'}^2+4\mu_{{\rm B}}^2H^2\omega_n^2\Bigl)
\end{eqnarray}

Close inspection of the above equations reveals that they have the following structure:
\begin{eqnarray}
M_{\bi k} &=&\sum_n \sum_{{\bi k'}}V_{\bi k k'}^{SDW}
\Bigl\{M_{{\bi k'}} \bigl\{...\bigr\}+\delta_{\bi k'} \Delta_{\bi k'}
\Pi_{\bi k'}\bigl\{...\bigr\}\Bigr\} \nonumber \\ \Delta_{\bi k}
&=&\sum_n \sum_{{\bi k'}}V_{\bi k k'}^{dSC} \Bigl\{\Delta_{{\bi k'}}
\bigl\{...\bigr\}+\delta_{\bi k'} M_{\bi k'}\Pi_{\bi k'}
\bigl\{...\bigr\}\Bigr\} \nonumber \\ \Pi_{\bi k} &=&\sum_n
\sum_{{\bi k'}}V_{\bi k k'}^{Qtr} \Bigl\{\Pi_{{\bi k'}}
\bigl\{...\bigr\}+\delta_{\bi k'} M_{\bi k'} \Delta_{\bi
k'}\bigl\{...\bigr\}\Bigr\} \end{eqnarray}
On the right hand side of each of the gap equations,
there are terms which {\it are not proportional to the gap of the left hand side}.
When there is particle-hole asymmetry, then if two of the order parameters are non-zero,
zero is not a trivial self-consistent solution for the third order parameter
which has to be non-zero as well. Furthermore, the induced term by p-h
asymmetry is even in frequency, so we expect it not to vanish after the
summation on the matsubara frequencies. Therefore,
in the presence of both SDW and d-wave singlet SC orderings,
\emph{particle-hole asymmetry would imply the
presence of a staggered $\pi$-triplet SC component}.
Similar arguments are presented in \cite{Kyung} in a slightly different context.
Here we stress out that {\it partcle-hole asymmetry}
or the $\delta_{\bi{k}}$ term does the mixing.

The summations on the matsubara frequencies are done analytically
and the self-consistent gap equations on the real axis were solved
numerically to illustrate the mixing of the three order parameters.

\section{Numerical Results-Discussion}

We assume a tight-binding dispersion relation on a two dimensional
square lattice up to the n.n.neighbours. We set the nearest
neighbors hopping term $t_1=1.0$ and vary the n.n term $t_2$ which
is the relevant parameter for particle-hole asymmetry or doping. The
value of $t_1$ sets the energy scale in our calculations. The
effective potentials are generally decomposed into a momentum
dependent part, which includes any proper form factos, and a
constant part which sets the amplitude of the coupling strength. We
emphasize that in this study the SDW component is taken isotropic
and the SC components anisotropic assuming a d-wave pairing. As an
interesting example, we report our results for the case when the
above OPs have potentials of equal amplitude that cause a moderate
coupling, favoring a ground state of coexisting d-wave SC and SDW
components. Such a situation is realized when
$V^{SDW}=V^{dSC}=V^{Qtr}=3$. This scenario resembles the case where
$T_c>T_N$ in\cite{Machida2}.
\begin{figure}[htbp]\begin{center}
\includegraphics[width=0.45\textwidth]{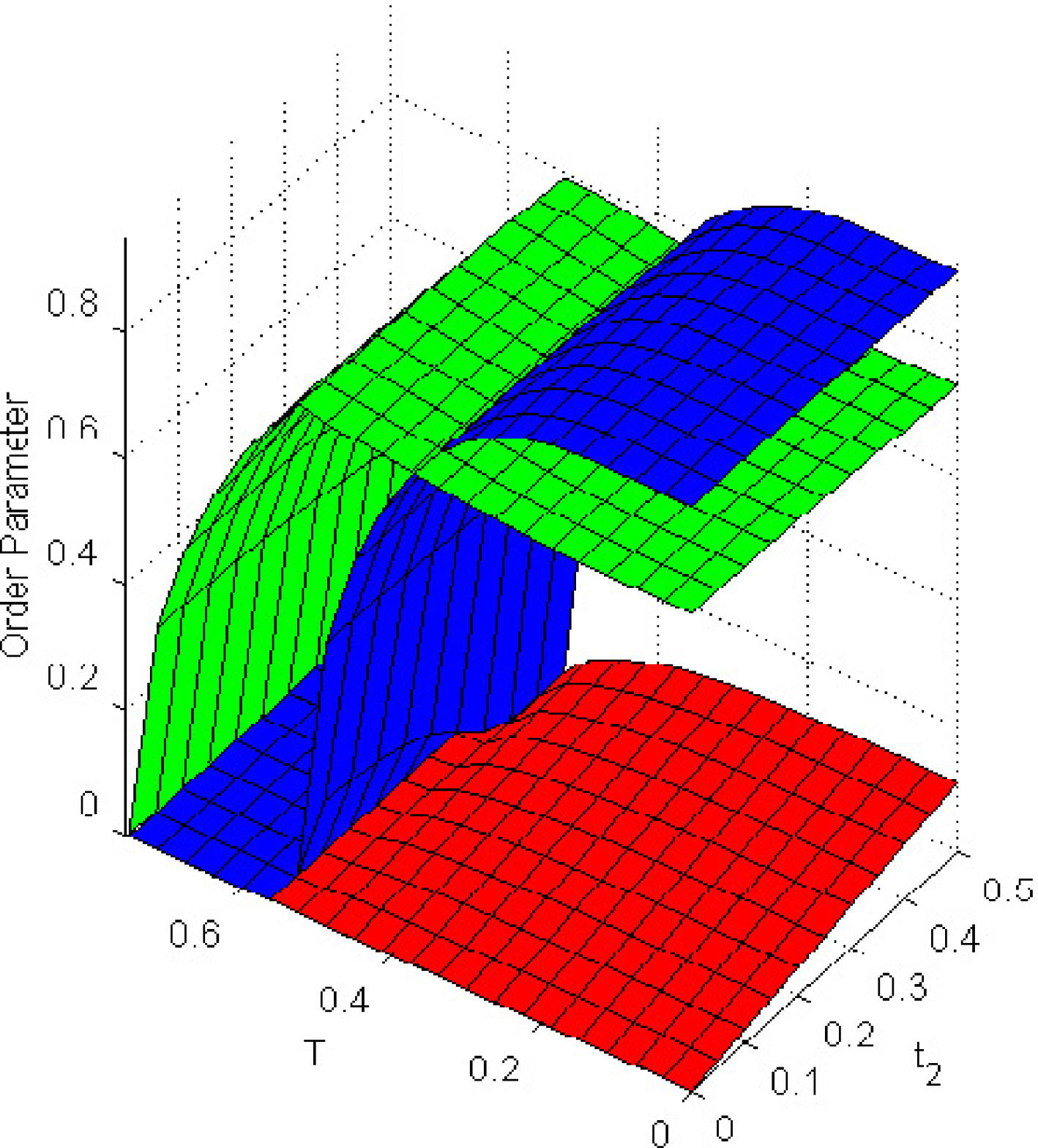}
\includegraphics[height=0.45\textwidth,width=0.45\textwidth]{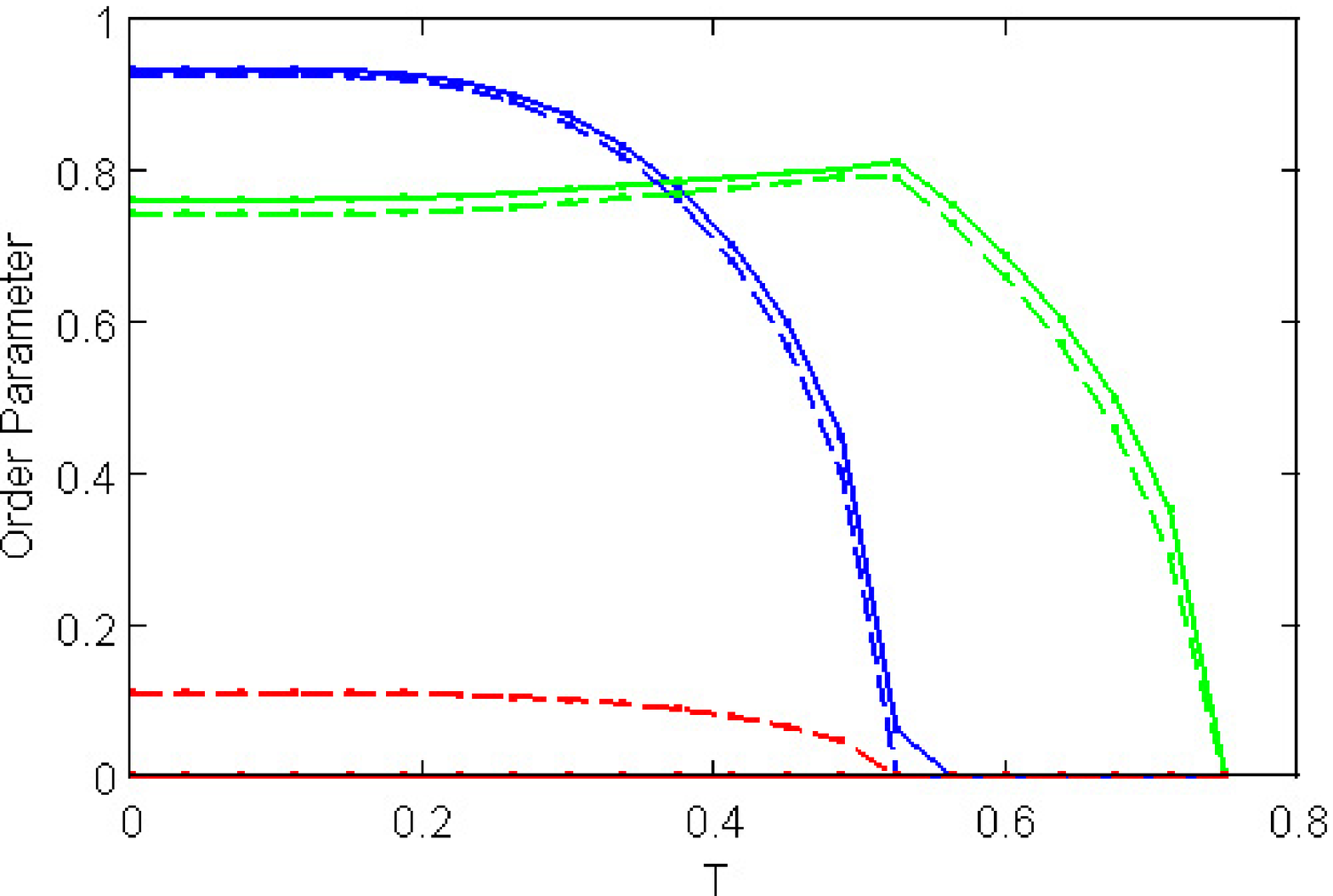}
\caption{(a) An example of how p-h asymmetry induces a
staggered\\$\pi$-triplet SC component (red). No external field is
assumed here ($\mu_{{\rm B}}H=0$) We note that when we have perfect
nesting (i.e. $t_2=0$) there is at low temperatures coexistence of
SDW with d-wave singlet SC (blue and green respectively) and above a
given temperature we have a transition to a state where only d-wave
singlet SC is present. We note that the staggered $\pi-triplet$ SC
component (red) appears for $t_2\neq 0$. In fact with $t_2\neq 0$ it
is impossible to obtain any of the two order parameters coexisting
without the third one whatever the choice o the effective potentials
or the exact electronic dispersions.
\\(b) We present two different 2-D cuts on the previous 3-D figure,
one for $t_2=0$ (full lines) and the other one for $t_2=0.5$ (dashed
lines). Note that while particle-hole asymmetry has negligible
influence on the temperature behavior, it induces the staggered
$\pi-triplet$ component almost at the same temperature at which the
dSC-SDW coexistence would arise. \label{p-has}}
\end{center}\end{figure}
Notice that we have chosen blue,green and red colouring to discern the SDW,d-wave singlet and $\pi$-triplet OPs respectively. In Figure \ref{p-has} we show that at the ground state and at
half-filling this system orders in an antiferromagnetic nodal
superconductor (d-wave). For a finite $\delta_{\bi{k}}$ (e-doping),
a $\pi$-triplet SC component is imposed and we end up with an
antiferromagnetic superconductor in both singlet and triplet
channels. Here, $t_2$ is sufficiently large to induce deviations
from perfect nesting, but not enough to destroy the SDW component.
We readily verify that \textit{any} finite $t_2$ \textit{mixes} the
considered OPs. We will present data for $t_2=0.5$ from now on as
the phenomena we describe are more intense. Our qualitiative results
do not differ at all for lower values of $t_2$.

\begin{figure}[htbp]\begin{center}
\includegraphics[height=0.4\textwidth,width=0.45\textwidth]{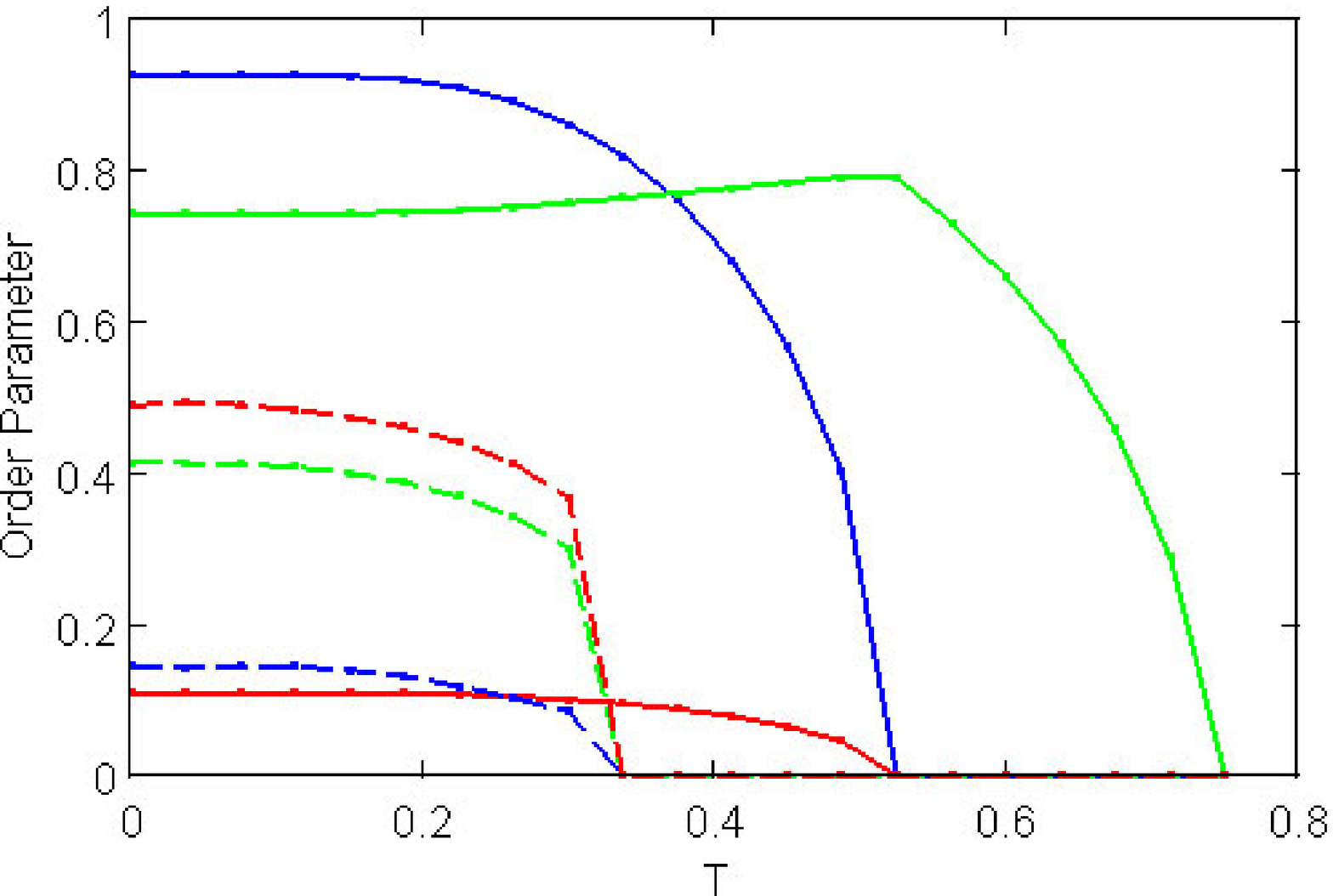}
\includegraphics[height=0.4\textwidth,width=0.45\textwidth]{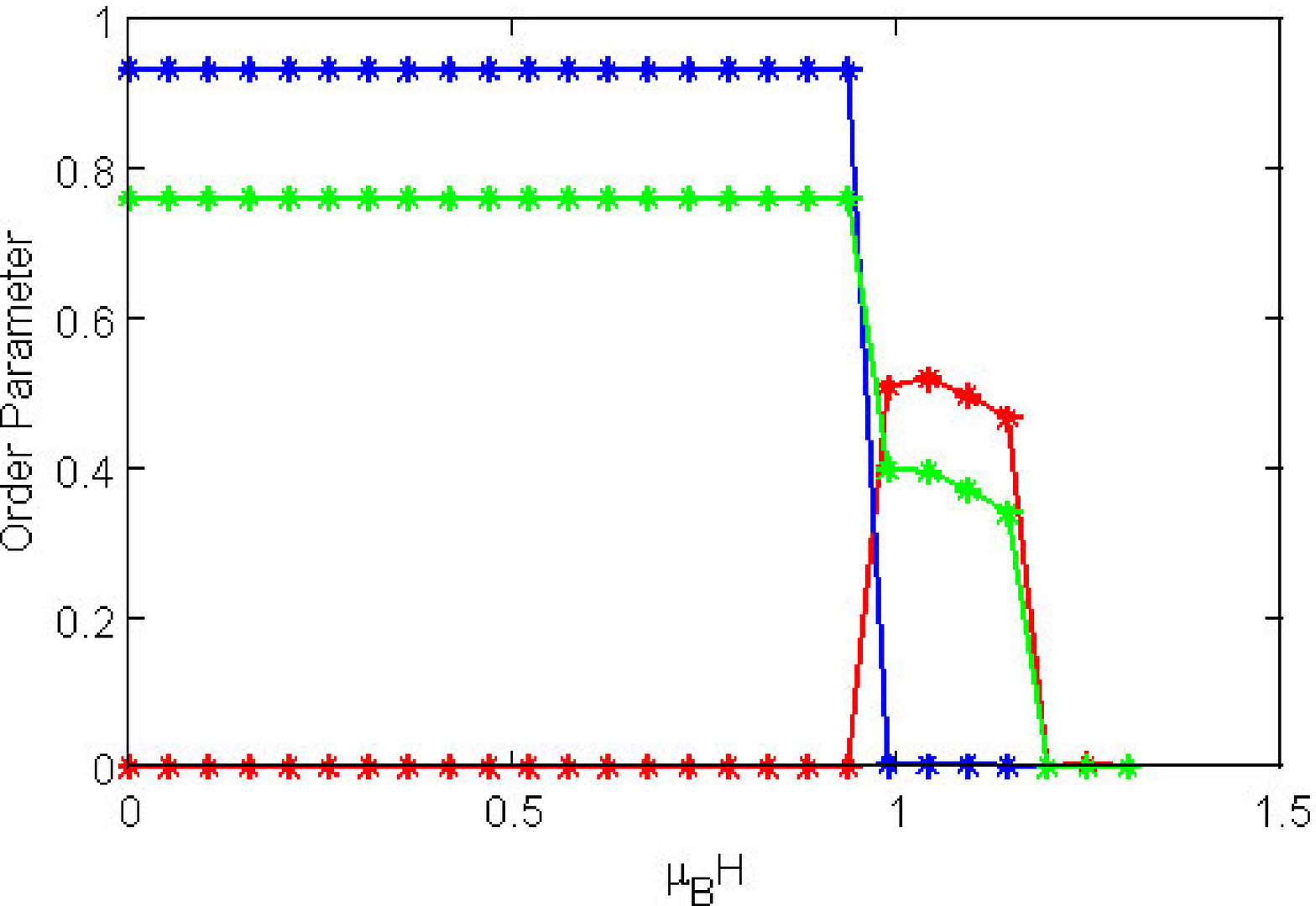}
\caption{(a) We illustrate the influence of a uniform magnetic field
in a particle-hole asymmetric system ($t_2=0.5$). With no external
field (full lines) the system exhibits the coexistence of the three
OPs below a given temperature. At a finite external field (dashed
lines), the three order parameters arise simultaneously in a
\textit{first order transition at the same temperature}. The field
reduces drastically the SDW (blue) producing a dominant
$\pi$-triplet SC component (red) in the low-T regime. (b) The field
dependence of the OPs at the ground state ($T=0$) and half filling ($t_2=0$). Varying the
magnetic field we induce a first order transition from the SDW+dSC
state to a mixed singlet-triplet SC order. This transition exhibits
many similarities to the transitions to a FFLO state.\label{infmag}}
\end{center}\end{figure}
In Figure \ref{infmag} we see that at low-T, applying a magnetic
field (here, $\mu_BH=0.78)$ we induce transitions essentially from the SDW + d-SC phase to
the d-SC + $\pi$-SC phase. These exhibit the characteristics of the
Fulde Ferrel transitions yet they are qualitatively different.It is
quite remarkable that the transition from the SDW + d-SC phase to
the d-SC + $\pi$-SC is first order and would be hard to
differentiate from a conventional FFLO state. Although our high
field d-SC + $\pi$-SC state has some similarity with the FFLO state
because the dominant SC order parameter is staggered (i.e. there is
a momentum modulation of the superfluid density), there are
fundamental differences between our d-SC + $\pi$-SC state and the
FFLO state. Firstly, the FFLO state is considered to be a singlet
state, whereas our staggered state is triplet. Moreover, in the FFLO
state there are regions of the FS which are superconducting and
regions of the FS which are not gapped et al. (they are normal). In
our case, all regions of the FS are gapped. This is in fact a
transition to a novel SC state in which singlet and triplet
staggered components coexist. Note that this last coexistence do not
requires the presence of particle-hole asymmetry. If the system was
particle-hole asymmetric, the high field state would involve a weak
SDW component as well since as we have noticed it is impossible to
observe any pair of the above components without the third one.

\section{Conclusions}

In conclusion, we have shown that particle-hole asymmetry mixes
d-wave singlet SC with $\pi-triplet$ staggered SC and SDW. We may
either observe one of them, or else all three of them. As a result,
in any SDW superconductor there are both singlet and staggered
triplet superconducting components. This may be behind the unsettled
controversies about the parity of the oredr parameter in many of the
SDW superconducting systems. Moreover, we have shown that the
application of a uniform external magnetic field induces new
transitions that exhibit remarkable similarities with the
Fulde-Ferrel phases. We believe that our new SC states, where
basically d-wave singlet SC and staggered $\pi-triplet$ SC coexist
the later being dominant, may be in fact behind the signatures of
FFLO phases that are reported in Ce based heavy fermion compounds.

\ack This work has been supported by the European Union through the
STRP NMP4-CT-2005-517039 CoMePhS grant.

\end{document}